\documentstyle[12pt]{article} 
\begin{document}
\def\theequation{\arabic{section}.\arabic{equation}}
\newcommand{\be}{\begin{equation}} 
\newcommand{\ee}{\end{equation}}
\begin{titlepage} 
\title{Solving for the dynamics of the universe} 
\author{Valerio Faraoni \\ \\
{\small \it Research Group in General Relativity (RggR)}\\
{\small \it Universit\'{e} Libre de Bruxelles }\\
{\small \it Campus Plaine CP231, Blvd. du Triomphe}\\
{\small \it 1050 Bruxelles, Belgium}\\
{\small \it E--mail: vfaraoni@ulb.ac.be}
} \date{} 
\maketitle
\thispagestyle{empty} 
\vspace*{1truecm} 
\begin{abstract} 
A new method of solving the Einstein--Friedmann dynamical equations of a
spatially homogeneous and isotropic universe is presented. The
method is applicable when the equation of state of the material content 
assumes the form $P=\left( \gamma -1 \right) \rho $, $ \gamma = $~constant. 
The solution for the expansion factor is commonly given only for
$\gamma =0,1$ and $4/3$ when the curvature index is $K= \pm 1$. The proposed
procedure is valid for {\em general} $\gamma$ and $K$ and it
allows for ease of derivation of the solutions. This alternative method is
useful pedagogically to introduce basic cosmology.
\end{abstract} 
\vspace*{1truecm} 
\begin{center} 
To appear in {\em Am. J. Phys.}
\end{center} 
\end{titlepage} 

\section{Introduction} 

The Friedmann--Lemaitre--Robertson--Walker (FLRW) solutions to the Einstein
field equations of general relativity are a cornerstone in the development
of modern cosmology. The FLRW metric describes a spatially homogeneous and
isotropic universe satisfying the Copernican principle, and is the starting
point for studying dynamical models of the universe. To this end, one must
specify the nature of the matter which is the source of gravitation by 
assigning the corresponding equation of state, which varies during different
epochs of the history of the universe.

The FLRW metric is given, in comoving coordinates $\left( t,r, \theta , 
\varphi \right)$, by the line element
\be  \label{1}
ds^2=-dt^2+a^2(t) \left[  dr^2 +f^2(r) \left( d\theta^2+\sin^2 \theta 
d\varphi^2 \right) \right] \; ,
\ee
where the three possible elementary topologies are classified according 
to a normalized curvature index $K$ which assumes the values $0, \pm 
1$, and 
\be  \label{2}
f(r)=\sinh r \;\;\;\;\;\;\;\;\;\;\;\;\;\;\;\;\;(K=-1)
\ee
for the open universe,
\be  \label{3}
f(r)= r \;\;\;\;\;\;\;\;\;\;\;\;\;\;\;\;\;(K=0)
\ee
for the critical universe, and
\be  \label{4}
f(r)= \sin r \;\;\;\;\;\;\;\;\;\;\;\;\;\;\;\;\;(K=+1)
\ee
for the closed universe. Other topologies are possible (see e.g. Ref.~1,
p.~725). 
The function $a(t)$ of the comoving time $t$  (the ``scale factor'') is 
determined by the Einstein--Friedmann dynamical equations$^{2-7}$
\be   \label{EF1}
\frac{\ddot{a}}{a}=-\, \frac{4\pi G}{3} \left( \rho +3P \right) \; ,
\ee
\be   \label{EF2}
\left( \frac{\dot{a}}{a} \right)^2= \frac{8\pi G \rho}{3} - \frac{K}{a^2} \; ,
\ee
where $\rho$ and $P$ are, respectively, the energy density and pressure 
of the material content of the universe, which is assumed to be a perfect 
fluid. An overdot
denotes differentiation with respect to the comoving time $t$, $G$ is 
Newton's constant
and units are used in which the speed of light in vacuum assumes the 
value unity. It is
further assumed that the cosmological constant vanishes.
 
One can solve the Einstein--Friedmann equations (\ref{EF1}) and (\ref{EF2}) 
once a barotropic
equation of state $P=P( \rho )$ is given. In many situations of physical 
interest, the equation of state assumes the form
\be   \label{7}
P=\left( \gamma -1 \right) \rho \; , \;\;\;\;\;\;\;\;\;\;\; \gamma 
=\mbox{constant}\; . 
\ee
This assumption reproduces important epochs in the history of the 
universe. For $\gamma
=1$ one obtains the ``dust'' equation of state $P=0$ of the 
matter--dominated epoch; for
$\gamma =4/3$, the radiation equation of state $P=\rho/3$ of the 
radiation--dominated era;
for $\gamma =0$, the vacuum equation of state $P=-\rho$ of inflation. For 
$\gamma=2/3$,
the curvature--dominated coasting universe with $a(t)=a_0 t$ is obtained.

When the universe is spatially flat ($K=0$), the integration of the 
dynamical equations (\ref{EF1}), (\ref{EF2}) with the assumption (\ref{7}) is
straightforward$^2$ for any value of the constant $\gamma$:
\be \label{8}
a(t)=a_0 t^{\frac{2}{3\gamma}}  \;\;\;\;\;\;\;\;\;\;\;\;( \gamma \neq 0) \; ,
\ee
\be  \label{9}
a(t)=a_0 \mbox{e}^{Ht}\; , \;\;\;\;\;\; \dot{H}=0  
\;\;\;\;\;\;\;\;\;\;\;\;( \gamma = 0) \; .
\ee
However, for $K=\pm 1$, the solution is given (explicitly or in 
parametric form) only for the special values $ \gamma =1, 4/3$ in old and
recent textbooks$^{1-7}$.

It is actually not difficult to derive a solution
for any nonzero value of the constant $\gamma$ using a standard
procedure$^3$ which is general, i.e. it does not
depend upon the assumption $P=\left( \gamma -1 \right) \rho $. Section~2
presents a summary of the usual method for deriving the scale factor $a(t)$.
An alternative method, which consists in reducing the Einstein--Friedmann
equations to a Riccati equation, and is applicable when Eq.~(\ref{7})
holds, is explained in Sec.~3.
From the mathematical point of view, the latter method avoids the consideration of
the energy conservation  equation 
\be  \label{conservation}
\dot{\rho} +3\left( P+\rho \right) \, \frac{\dot{a}}{a}=0 
\ee
and the calculation of an indefinite integral; rather, one solves a 
nonlinear Riccati equation. The alternative approach is more
direct than the standard one and the gain in clarity of exposition makes it
more suitable for an introductory cosmology course. Section~4 presents a brief 
discussion of the equation of state $P=\left( \gamma -1 \right) \rho$ and the
conclusions.

\section{The standard derivation of the scale factor} 

\setcounter{equation}{0}

The standard method to obtain the scale factor $a$ of the universe 
proceeds as follows$^3$. The energy
conservation
equation (\ref{conservation}) yields
\be   \label{22}
3\ln a=-\int \frac{d\rho}{P+\rho} +\mbox{constant}
\ee
for $\gamma \neq 0$. Upon use of the conformal time $\eta$ defined by
\be  \label{23}
dt=a( \eta ) d\eta \; ,
\ee
the Einstein--Friedmann equations (\ref{EF1}), (\ref{EF2}) yield
\be  \label{24}
\eta =\pm \int \frac{da}{a\sqrt{\frac{8\pi G}{3} \rho a^2 -K}} \; .
\ee
In order to obtain $a( \eta )$, one prescribes the equation of state $P=P( 
\rho) $, solves Eq.~(\ref{22}) and inverts it obtaining $\rho=\rho (a) $. Further 
substitution into Eq.~(\ref{24}) and inversion provide $a=a( \eta )$. Integration
of Eq.~(\ref{23}) provides the comoving time $t( \eta )$ as a function of conformal
time.  The scale factor
is then expressed in parametric form $\left( a( \eta ), t( \eta ) 
\right)$. Sometimes it
is possible to eliminate the parametric dependence on $\eta$ and obtain 
the expansion factor as an explicit function $a(t) $ of comoving time.

The method is quite general; as a particular case, it can be applied when the 
equation of state is of the form $P=\left( \gamma -1 \right) \rho $ with
constant nonvanishing $\gamma $. Equations (\ref{22}) and 
(\ref{24}) then yield
\be  \label{25}
a^3 \rho^{1/\gamma}=\mbox{constant} \; ,
\ee
\be  \label{26}
\eta =\pm \int \frac{ da}{ a \sqrt{\frac{8\pi G}{3} C_1 a^{2-3\gamma}-K}}
\ee
for $\gamma \neq 0, 2/3$, where $C_1$ is an integration constant. By 
introducing the variable
\be  \label{27}
x=\left( \frac{8\pi G C_1}{3} \right)^{\frac{1}{2-3\gamma}} a \; ,
\ee
and using 
\be  \label{28}
\int \frac{dx}{ x\sqrt{x^n+1}}=\frac{1}{n} \ln \left(
\frac{\sqrt{x^n+1}-1}{\sqrt{x^n+1}+1} \right) \; ,
\ee
\be  \label{29}
\int \frac{dx}{ x\sqrt{x^n-1}}=\frac{2}{n}\, \mbox{arcsec} \left(
x^{n/2} \right) \; ,
\ee
one integrates and inverts Eq.~(\ref{26}) to obtain
\be  \label{210}
a( \eta )= a_0 \sinh^{1/c} ( c\eta ) \; ,
\ee
\be  \label{211}
t( \eta )= a_0 \int_0^{\eta} d\eta ' \sinh^{1/c} ( c\eta ' ) \; ,
\ee
for $K=-1$. $a_0$ and $c$ are constants, with
\be    \label{212}
c=\frac{3}{2} \, \gamma -1 \; ,
\ee
and the boundary condition $a \left( \eta=0 \right) =0$ has been imposed.

Similarly, for $K=+1$, one obtains
\be  \label{213}
a( \eta )= a_0 \left[ \cos \left( c\eta +d \right) \right]^{1/c} \; ,
\ee
\be   \label{214}
t( \eta )= a_0 \int_0^{\eta} d\eta ' \left[ \cos \left( c\eta '+d \right) 
\right]^{1/c} \; .
\ee
For $\gamma=2/3$ and $K=-1$ one obtains a curvature--dominated universe for
which Eq.~(\ref{EF2}) is approximated by $ \left( \dot{a}/a \right)^2
\simeq -K/a^2$. In this case, Eq.~(\ref{24}) yields 
$a=a_0 \exp( \beta \eta)$ ($\beta=$~constant), and $t=t_0$e$^{\eta 
\beta}$ gives $a=a_0t$. It is easier to obtain this form of the scale factor
directly  from  Eq.~(\ref{EF1}), which reduces to $\dot{a}=0$.

The solutions (\ref{210}), (\ref{211}) and (\ref{213}), (\ref{214}) for 
the scale factor are presented in
the textbooks only for the special values $1$ and $4/3$ of the constant 
$\gamma$. For $\gamma =4/3$ one eliminates the parameter $\eta$ to obtain
\be
a(t)= a_0 \left[ 1-\left( 1-\frac{t}{t_0} \right)^2 \right]^{1/2} 
\ee
for $K=+1$, and
\be
a(t)= a_0 \left[ \left( 1+\frac{t}{t_0} \right)^2 -1 \right]^{1/2} 
\ee
for $K=-1$.

The standard solution method of the Einstein--Friedmann equations has the
virtue of being general; it does not rely upon the assumption
(\ref{7}). However, the need to invert Eqs.~(\ref{22}) and (\ref{24})
and to compute the indefinite integrals (\ref{27}), (\ref{28}) detracts from
the elegance and clarity that is possible when the ratio $P/\rho $ is
constant. The latter condition is satisfied in many physically important
situations.

\section{An alternative method} 

\setcounter{equation}{0}

There is an alternative procedure to derive the scale factor for a general
value of $\gamma$ when the equation of state of the universe's material
content is given by $P=\left( \gamma -1 \right)  \rho$ with $\gamma
=$~constant, which covers many cases of physical interest. Being
straightforward, this new method is valuable for pedagogical purposes and
proceeds as follows: Eqs.~(\ref{EF1}), (\ref{EF2}) and (\ref{7}) yield \be
\label{215} \frac{\ddot{a}}{a}+c\left( \frac{\dot{a}}{a}
\right)^2+\frac{cK}{a^2} =0 \; , \ee with $c$ given by Eq.~(\ref{212}).
For $K=0$, Eq.~(\ref{215}) is immediately integrated to give
Eqs.~(\ref{8}), (\ref{9}). For $K=\pm 1$, Eq.~(\ref{215}) is rewritten as
\be \label{216} \frac{a''}{a}+(c-1) \left( \frac{a'}{a} \right)^2+cK =0 \;
, \ee by making use of the conformal time $\eta$, and where a prime
denotes differentiation with respect to $\eta$. By employing the variable
\be \label{217} 
u \equiv \frac{a'}{a} \; , 
\ee 
Eq.~(\ref{216}) becomes 
\be
\label{218} 
u'+cu^2+Kc=0 \; , 
\ee 
which is a Riccati equation. The Riccati
equation, which has the general form 
\be 
\frac{dy}{dx}=a(x) y^2 +b(x) y +c (x) 
\ee
where $y=y(x) $, has been the subject of many studies in the theory of
ordinary differential equations, and can be solved explicitly$^{8,9}$.
The
solution is found by introducing the variable $w$ defined by 
\be  \label{219}
u = \frac{1}{c}\, \frac{w'}{w} \; ,
\ee
which changes Eq.~(\ref{218}) to
\be  \label{220}
w''+Kc^2w=0 \; ,
\ee
the solution of which is trivial.
For $K=+1$, one finds the solutions (\ref{213}), (\ref{214}), while for 
$K=-1$, one recovers (\ref{210}), (\ref{211}).

The alternative method is applicable when the equation of
state assumes the form $P=\left( \gamma -1 \right) \rho $ and 
the cosmological constant vanishes. With these conditions satisfied,
the alternative solution procedure is more direct than the general
method.
It is now appropriate to comment on the equation of state $P=\left( \gamma 
-1 \right) \rho$, $\gamma=$~constant, which reproduces several situations of
significant  
physical interest.

\section{Discussion}

\setcounter{equation}{0}

The assumption that the equation of state is of the form $P=\left( \gamma -1 
\right) \rho$ with constant $\gamma$ is justified in many important
situations which describe the standard big--bang cosmology. However,
it is important to realize that Eq.~(\ref{7}) is a strong assumption and by
no means yields the most general solution for the scale factor of a FLRW 
universe. To make a physically interesting example, consider the 
inflationary  (i.e. $\ddot{a}>0$) epoch of the early universe in
the $K=0$ case. Many inflationary scenarios are known$^{10}$, 
corresponding to different concave shapes of the scale factor $a(t)$; 
the assumption (\ref{7}) allows the solutions 
\be
a(t)=a_0 t^{\frac{2}{3\gamma}}
\ee
(``power--law inflation'') for $0<\gamma < 2/3$, and 
\be
a=a_0{\mbox e}^{Ht} \; , \;\;\;\;\;\; \dot{H}=0 
\ee
for $\gamma =0$. Vice--versa, using the dynamical equations (\ref{EF1}), 
(\ref{EF2}), it is straightforward to prove that the latter solutions
imply $P/\rho=$~constant. The assumption (\ref{7}) reproduces {\em only}  
exponential expansion (the prototype of inflation) and power--law inflation.
All the other inflationary scenarios correspond to a $\gamma (t ) $ which
changes with time during inflation. A time--dependent $\gamma (t)$ can
also be used to describe a non--interacting mixture of dust and radiation;
however the method of solution of the Einstein--Friedmann equations 
(\ref{EF1}), (\ref{EF2}) presented in Sec.~3 applies only when $\gamma$ is
constant, and when the cosmological constant vanishes. If the cosmological
constant is nonzero, Eq.~(\ref{215}) does not reduce to
a Riccati equation.

The limitations of the assumption $\gamma=$~constant are thus made clear:
while this new approach to the Einstein--Friedmann equations does not
replace the standard approach, it is more direct and is  preferable in an
introduction  to cosmology. Its value lies in the ease of demonstration of
the solutions, which is crucial for students to grasp the basic concepts of
cosmology.

\section*{Acknowledgments}

The author thanks L. Niwa for reviewing the manuscript and two anonymous
referees for useful comments.

\vskip1.5truecm

\noindent $^1$ C.W. Misner, K.S. Thorne and J.A. Wheeler, {\em
Gravitation} (Freeman, San Francisco, 1973), pp.~733--742. \\
$^2$ S. Weinberg, {\em Gravitation and Cosmology} (J.
Wiley \& Sons, New York, 1972), pp.~475--491.
\\$^3$ L.D. Landau and E.M. Lifschitz, {\em The Classical Theory of 
Fields} (Pergamon Press, Oxford, 1989), pp.~363--367.
\\$^4$ R.M. Wald, {\em General Relativity} (Chicago University Press, 
Chicago, 1984), chap.~5.
\\$^5$ E.W. Kolb and M.S. Turner, {\em The Early Universe} 
(Addison--Wesley, Mass., 1990), pp.~58--60.
\\$^6$ T. Padmanabhan, {\em Cosmology and Astrophysics Through
Problems} (Cambridge University Press, Cambridge, 1996), pp.~79--89.
\\$^7$ R. D'Inverno, {\em Introducing Einstein's Relativity} (Clarendon
Press, Oxford, 1992), pp.~334--344.
\\$^8$ E. Hille, {\em Lectures on Ordinary Differential equations}
(Addison--Wesley, Reading, Mass., 1969), pp.~273--288.
\\$^9$ E.L. Ince, {\em Ordinary Differential Equations} (Dover, New York,
1944), pp.~23--25.
\\$^{10}$ A.R. Liddle and D.H. Lyth, {\em The Cold Dark Matter 
Density Perturbation}, Physics Reports {\bf 231}, 1--105 (1993).

\end{document}